\documentclass[twocolumn,showpacs,amsmath,amssymb,superscriptaddress]{revtex4}
\usepackage{graphicx}
\usepackage{dcolumn}
\usepackage{bm}
\usepackage[dvips]{epsfig}

\newcommand{\ix}[1]{{\raisebox{0pt}[0pt][0pt]{$\scriptstyle #1$}}}
\begin{document}
\title{Complete Integrability of Geodesic Motion in General Kerr-NUT-AdS
Spacetimes}

\affiliation{Theoretical Physics Institute, University of Alberta, Edmonton,
Alberta, Canada T6G 2G7}

\affiliation{JLR Engineering, 111 SE Everett Mall Way, E-201, Everett, WA
98208-3236, USA}

\affiliation{Institute of Theoretical Physics, Charles University, V
Hole\v{s}ovi\v{c}k\'ach 2, Prague, Czech Republic}

\author{Don N. Page}

\email{don@phys.ualberta.ca}

\affiliation{Theoretical Physics Institute, University of Alberta, Edmonton,
Alberta, Canada T6G 2G7}

\author{David Kubiz\v n\'ak}

\email{kubiznak@phys.ualberta.ca}

\affiliation{Theoretical Physics Institute, University of Alberta, Edmonton,
Alberta, Canada T6G 2G7}

\affiliation{Institute of Theoretical Physics, Charles University, V
Hole\v{s}ovi\v{c}k\'ach 2, Prague, Czech Republic}

\author{Muraari Vasudevan}

\email{mvasudev@phys.ualberta.ca}

\affiliation{Theoretical Physics Institute, University of Alberta, Edmonton,
Alberta, Canada T6G 2G7}

\affiliation{JLR Engineering, 111 SE Everett Mall Way, E-201, Everett, WA
98208-3236, USA}

\author{Pavel Krtou\v{s}}

\email{Pavel.Krtous@mff.cuni.cz}

\affiliation{Institute of Theoretical Physics, Charles University, V
Hole\v{s}ovi\v{c}k\'ach 2, Prague, Czech Republic}

\date{2006 Nov. 13}

\begin{abstract}

We explicitly exhibit $n-1$ constants of motion for geodesics in the general
$D$-dimensional Kerr-NUT-AdS rotating black hole spacetime, arising from
contractions of even powers of the $2$-form obtained by contracting the
geodesic velocity with the dual of the contraction of the velocity with the
$(D-2)$-dimensional Killing-Yano tensor.  These constants of motion are
functionally independent of each other and of the $D-n+1$ constants of motion
that arise from the metric and the $D-n = [(D+1)/2]$ Killing vectors, making a
total of $D$ independent constants of motion in all dimensions $D$.  The
Poisson brackets of all pairs of these $D$ constants are zero, so geodesic
motion in these spacetimes is completely integrable.

\end{abstract}

\pacs{04.70.Bw, 04.50.+h, 04.20.Jb \hfill  Alberta-Thy-13-06}

\maketitle

With motivations especially from string theory, many people have shown much
recent interest in black hole metrics in higher dimensions.  Nonrotating black
hole metrics in higher dimensions were first given in 1963 by Tangherlini
\cite{Tang}.  In 1986 Myers and Perry \cite{MP} generalized the 1963 Kerr
metric \cite{Kerr} for a $4$-dimensional rotating black hole to all higher
dimensions $D$. In 1968 Carter \cite{Carter} added a cosmological constant to
get a $4$-dimensional rotating Kerr-de Sitter metric.  In 1998 Hawking, Hunter,
and Taylor-Robinson \cite{HHT} found the general $5$-dimensional extension of
this metric, and in 2004 Gibbons, L\"u, Page, and Pope \cite{GLPP1,GLPP2}
discovered the general Kerr-de Sitter metrics in all higher dimensions.  Just
this year Chen, L\"u, and Pope \cite{CLP} were able to add a NUT \cite{NUT}
parameter to get the general Kerr-NUT-AdS metrics in all dimensions, which they
presented in an especially simple form, analogous to the
Pleba\'nski-Demia\'{n}ski \cite{PD} $4$-dimensional generalization of Carter's
Kerr-de Sitter metric.

With $n=[D/2]$, $\varepsilon=D-2n$, $m=n-1+\varepsilon$, Latin indices running
over $1$ through $D = 2n + \varepsilon$, and Greek indices running over $1$
through $n$, these Kerr-NUT-AdS metrics \cite{CLP} that solve the Einstein
equation $R_{ab} = - (D-1) g^2 g_{ab}$ may, after suitable analytic
continuations, be written in the orthonormal form (cf.\ \cite{KF})
\begin{eqnarray}\label{metrics}
ds^2 \equiv \mathbf{g}
     &=& \sum_{a=1}^D \sum_{b=1}^D \delta_{ab}\mathbf{e}^a \mathbf{e}^b
         \nonumber\\
     &=& \sum_{\mu=1}^n 
          (\mathbf{e}^\mu \mathbf{e}^\mu + \mathbf{E}^\mu \mathbf{E}^\mu)
       + \varepsilon \mathbf{\omega\omega} \, ,
\end{eqnarray}
where the orthonormal basis one-forms are
\begin{eqnarray}\label{one-forms}           
\mathbf{e}^\mu &=& Q_{\mu}^{-1/2} dx_{\mu}, \nonumber\\ 
\mathbf{e}^{n+\mu} = \mathbf{E}^\mu &=& Q_{\mu}^{1/2}
 \sum_{k=0}^{n-1}A_{\mu}^{(k)}d\psi_k, \nonumber\\
\mathbf{e}^{2n+1} = \mathbf{\omega} &=& (-c/A^{(n)})^{1/2}
\sum_{k=0}^nA^{(k)}d\psi_k,
\end{eqnarray}
and where
\begin{eqnarray}\label{co}
Q_{\mu}&=&\frac{X_{\mu}}{U_{\mu}},\quad
U_{\mu}=\prod_{\nu=1}^{\prime\,n}(x_{\nu}^2-x_{\mu}^2),\quad c=\prod_{k=1}^m a_k^2, \nonumber\\
X_{\mu}&=&(-1)^{\varepsilon}\frac{g^2x_{\mu}^2-1}{x_{\mu}^{2\varepsilon}}\prod_{k=1}^{m}(a_k^2-x_{\mu}^2)
+2M_{\mu}(-x_{\mu})^{(1-\varepsilon)},\nonumber\\
A_{\mu}^{(k)}&=&\!\!\!\!\!\sum_{\nu_1<\dots<\nu_k}^{\prime}\!\!\!\!\!x^2_{\nu_1}\dots x^2_{\nu_k},\
A^{(k)}=\!\!\!\!\!\sum_{\nu_1<\dots<\nu_k}\!\!\!\!\!x^2_{\nu_1}\dots x^2_{\nu_k}.
\end{eqnarray}
Primes on the sum and product symbols mean that the index $\nu=\mu$ is omitted. 
The $a_k$ are angular momentum parameters, and the $M_{\mu}$ are mass and NUT
parameters. 

The inverse Kerr-NUT-AdS metric has the form
\begin{eqnarray}\label{inverse}
\left(\frac{\partial}{\partial s}\right)^2
       &=& \sum_{a=1}^D \sum_{b=1}^D \delta^{ab} \mathbf{e}_a \mathbf{e}_b
       \nonumber\\
       &=& \sum_{\mu=1}^n
          (\mathbf{e}_\mu \mathbf{e}_\mu + \mathbf{E}_\mu \mathbf{E}_\mu)
       + \varepsilon \mathbf{E E},
\end{eqnarray}
where the orthonormal basis vectors are
\begin{eqnarray}\label{basisvectors}
\mathbf{e}_\mu &=& Q_{\mu}^{1/2} \frac{\partial}{\partial x_{\mu}}, \nonumber\\
\mathbf{e}_{n+\mu} = \mathbf{E}_\mu
&=& Q_{\mu}^{-1/2}U_{\mu}^{-1}
\sum_{k=0}^{m}(-1)^{n-1-k} x_{\mu}^{2(n-1-k)}\frac{\partial}{\partial \psi_{k}},
\nonumber\\
\mathbf{e}_{2n+1} = \mathbf{E} &=& (-cA^{(n)})^{-1/2}
\frac{\partial}{\partial \psi_n}.
\end{eqnarray}

Kubiz\v n\'ak and Frolov \cite{KF} have shown that the Kerr-NUT-AdS metric
possesses a $(D-2)$-rank Killing-Yano tensor 
\begin{equation}\label{KY}
\mathbf{f} = \mathbf{*k},
\end{equation}
where the closed $2$-form $\mathbf{k}$ can easily be shown to be
\begin{equation}\label{k}
\mathbf{k} = \sum_{\mu=1}^n x_\mu \mathbf{e}^\mu \wedge \mathbf{E}^\mu.
\end{equation}

In the general case a Killing-Yano tensor \cite{Yano} of rank $p$ is a $p$-form
$\mathbf{f}$ that satisfies the equations 
\begin{equation}\label{Yeq}
f_{a_1\dots\ a_p}=f_{[a_1\dots a_p]},\quad 
f_{a_1\dots (a_p;a_{p+1})}=0.
\end{equation} 

Kubiz\v n\'ak and Frolov \cite{KF} then show how to use the conformal Killing
tensor
\begin{eqnarray}\label{CKT}
\mathbf{Q} &=& Q_{ab} \mathbf{e}^a \mathbf{e}^b
 = k_{ac} k_b^{\ c} \mathbf{e}^a \mathbf{e}^b \nonumber\\
 &=& \sum_{\mu=1}^n x_\mu^2
 (\mathbf{e}^\mu \mathbf{e}^\mu + \mathbf{E}^\mu \mathbf{E}^\mu)
\end{eqnarray}
to construct the $2$nd-rank Killing tensor that can easily be shown to be
\begin{equation}\label{1KT}
\mathbf{K}=\mathbf{Q}-\frac{1}{2}\, Q^{\ c}_{c}\mathbf{g}
= -\sum_{\mu=1}^n A^{(1)}_\mu
(\mathbf{e}^\mu \mathbf{e}^\mu + \mathbf{E}^\mu \mathbf{E}^\mu)
-\varepsilon A^{(1)}\omega\omega.
\end{equation}

A Killing tensor \cite{Stac,WP,Car2} of rank $r$ is a totally symmetric tensor
$\mathbf{K}$ that satisfies the equations
\begin{equation}\label{KillTens}
K_{a_1\dots a_r}=K_{(a_1\dots a_r)},\quad 
K_{(a_1\dots a_r; a_{r+1})}=0.
\end{equation} 

Geodesic motion gives conserved constants from contractions of one velocity
$\mathbf{u} = u^a \mathbf{e}_a$ with each of the $D-n = n + \varepsilon = m+1$
Killing vectors $\partial/\partial \psi_{k}$, from the contraction of two
velocities with the metric, and from contractions of velocities with any
Killing tensors present.  With one $2$nd-rank Killing tensor present that is
independent of the metric (which is always a Killing tensor), one thus has
$D+2-n$ constants of motion. For $n\leq 2$ or $D \leq 5$, this gives a full set
of $D$ constants to make the geodesic motion integrable.  However, for $D>5$,
it was not previously known how to find a full set of $D$ constants of geodesic
motion for the general Kerr-NUT-AdS metrics \cite{CLP}, or even for the general
Myers-Perry (MP) metrics \cite{MP} obtained by eliminating the NUT parameters
and the cosmological constant. For earlier work on geodesic motion and Killing
tensors in the MP, Kerr-(NUT)-AdS, and related metrics in higher dimensions,
see \cite{KF,Palm,FS1,FS2,CGLP,Vas1,Vas2,KL1,KL2,Vas3,Vas4,DKL,Vas5,CLP1,Dav1,
KF2,CCLP,Dav2}.

The point of the present letter is to show that one can obtain a full set of
$D$ independent constants in involution for geodesic motion in the general
Kerr-NUT-AdS metrics, thereby making this motion completely integrable.

Briefly, the demonstration uses the fact that when the velocity is contracted
with the Killing-Yano tensor of rank $D-2$, this gives a $(D-3)$-form that is
covariantly constant along each geodesic.  The dual of this $(D-3)$-form gives
a $3$-form that is also covariantly constant along each geodesic, as is the
$2$-form contraction of this $3$-form with the velocity.  This $2$-form has at
least $n-1$ nonzero complex-conjugate (pure imaginary) pairs of eigenvalues
that give $n-1$ constants of motion, which we can show \cite{KKPV} are
independent of each other and of the $D-n+1$ constants of motion obtainable
from the Killing vectors and the metric, with all of the Poisson brackets
between them vanishing.  Therefore, we have $D$ independent constants of motion
in involution for geodesics in the general Kerr-NUT-AdS metrics, making the
geodesics completely integrable (see, e.g., \cite{Arn,Koz}).

Let us write the resulting $2$-form as
\begin{equation}\label{2form}
\mathbf{F}=\mathbf{u\cdot(*(u\cdot(*k)))}
= \frac{1}{2} F_{ab} \,\mathbf{e}^a\! \wedge \mathbf{e}^b,
\end{equation}
with components
\begin{equation}\label{2formc}
F_{ab} = (k_{ab} u_c + k_{bc} u_a + k_{ca} u_b) u^c.
\end{equation}
Then since $\mathbf{F}$ is covariantly constant along geodesics, $u^c F_{ab;c}
=0$, the eigenvalues of $\mathbf{F}$ are constants of motion.  In particular,
the traces of even powers of the matrix form of $\mathbf{F}$ are constants. 
(The traces of odd powers are zero, because of the antisymmetry of
$\mathbf{F}$.)

Now let us give a formula for the new constants of motions ${C_j}$ that are
proportional to traces of the even powers of the matrix form of $\mathbf{F}$
and evaluate them explicitly for the $2$nd, $4$th, $6$th, and $8$th powers. 
For convenience, let us use matrix notation, in which $F$ is the antisymmetric
matrix with orthonormal components $F^a_{\ b}$, $K$ is the antisymmetric matrix
with components $k^a_{\ b}$ (not to be confused with a Killing tensor), $Q
\equiv -K^2$ is the symmetric matrix with components $Q^a_{\ b} = - k^a_{\ c}
k^c_{\ b}$, $W$ is the symmetric matrix with components $u^a u_b$, $w \equiv
{\rm Tr}(W) = u^c u_c$, $P \equiv I-W/w$ is the projection onto the hyperplane
orthogonal to the velocity, and $S \equiv -PKPKP$.  These matrices have the
properties that $P^2=P$ and $WK^{2j+1}W = 0$ for all nonnegative integers $j$. 
Then the component Eq. (\ref{2formc}) becomes the matrix equation
\begin{equation}\label{F}
F = wK - KW - W\!K = wPKP,
\end{equation}
whose negative square is the symmetric matrix
\begin{eqnarray}\label{F2}
-F^2 = w^2 S = w^2 PQP + w KWK.
\end{eqnarray}

One can now prove \cite{KKPV} that for all $j$,
\begin{equation}\label{Muraari}
{\rm Tr}(Q^j) + {\rm Tr}(S^j) = 2 {\rm Tr}[(QP)^j].
\end{equation}
Therefore, we get the constants of motion 
\begin{eqnarray}\label{cj}
C_{j} &\equiv& w^{-j}{\rm Tr}[(-F^2)^j] = w^{j} {\rm Tr}(S^j) \nonumber\\
&=& 2 {\rm Tr}[(wQ-QW)^j] - w^{j} {\rm Tr}(Q^j) \nonumber\\
&=& w^{j} {\rm Tr}(Q^j) - 2j\, w^{j-1} {\rm Tr}(Q^jW) \\
&&+\sum_{c=2}^{j}\sum_{\substack{l_1\leq\dots\leq l_c\\\sum_i l_i=j}}\!\!
  (-1)^c N^j_{l_1\dots l_j} w^{j-c} \prod_{i=1}^c{\rm Tr}(Q^{l_i}W)\;.\;\nonumber
\end{eqnarray}
where in the last expression the coefficients ${N^j_{l_1\dots l_j}}$ are some
positive combinatoric factors \cite{KKPV}.  We have used the fact that terms
${{\rm Tr}(Q^jWQ^kW\cdots)}$ factorize into ${({\rm Tr}Q^jW)({\rm
Tr}Q^kW)\cdots}$. Notice that all traces in the last line contain strictly
lower powers of $Q$ than $Q^j$.

To write the constants of motion in tensor notation, it is convenient to define
the scalars $Q^{(j)}$ that are the traces of the $j$th power of the matrix $Q$,
\begin{equation}\label{scalarQ}
Q^{(j)} \equiv {\rm Tr}(Q^j) = 2\sum_{\mu=1}^n x_\mu^{2j},
\end{equation}
and the symmetric covariant tensor components $Q^{(j)}_{ab}$ that form the
tensor $\mathbf{Q}^{(j)}$ corresponding to the $j$th power of the matrix $Q$,
\begin{equation}\label{tensorQ}
\mathbf{Q}^{(j)} = Q_{ab}^{(j)} \mathbf{e}^a \mathbf{e}^b
 = \sum_{\mu=1}^n x_\mu^{2j}
 (\mathbf{e}^\mu \mathbf{e}^\mu + \mathbf{E}^\mu \mathbf{E}^\mu).
\end{equation}
For example, $Q^{\ix{(1)}}\!=\!Q_c^{\ c}$, $Q_{ab}^{(1)}\!=\!Q_{ab}$,
$Q^{\ix{(2)}}\!=\!Q_c^{\ d} Q_d^{\ c}$, $Q_{ab}^{(2)}\!=\!Q_a^{\ c} Q^{}_{cb}$,
$Q^{\ix{(3)}}\!=\!Q_c^{\ d} Q_d^{\ e} Q_e^{\ c}$,
and $Q_{ab}^{(3)}\!=\!Q_a^{\ c} Q_c^{\ d} Q^{}_{db}$.

Then one can easily see that the ${C_j}$'s have the form
\begin{equation}\label{cisKus}
  C_j = K_{a_1\dots a_{2j}}u^{a_1}\dots u^{a_{2j}}
\end{equation}
for some symmetric tensors ${K_{a_1\dots a_{2j}}}$ formed from combinations of
the metric $g_{ab}$, $Q^{(j)}$, and the $Q^{(i)}_{ab}$'s for $i\leq j$. It can
be shown \cite{WP} that these are Killing tensors in the sense of
Eq.~\eqref{KillTens}.

In particular, we get
\begin{eqnarray}\label{1KTconst}
C_1 &=& w {\rm Tr}(Q) - 2{\rm Tr}(QW) \nonumber\\
&=& (Q^{\ix{(1)}}g_{ab} - 2 Q^{\ix{(1)}}_{ab}) u^a u^b = -2K_{ab} u^a u^b,
\end{eqnarray}
the constant from the previously-known 2nd-rank Killing tensor given in Eq.
(\ref{1KT}), and 
\begin{eqnarray}\label{2KTconst}
C_2 &=&w^2 {\rm Tr}(Q^2) - 4 w {\rm Tr}(Q^2W) + 2 [{\rm Tr}(QW)]^2 \nonumber\\
    &=&(Q^{\ix{(2)}}g^{}_{\ix{ab}}g^{}_{\ix{cd}} - 4 Q^{\ix{(2)}}_{\ix{ab}}g^{}_{\ix{cd}}
       + 2 Q^{\ix{(1)}}_{\ix{ab}}Q^{\ix{(1)}}_{\ix{cd}}) u^a u^b u^c u^d \nonumber\\
    &=& K_{abcd} u^a u^b u^c u^d \, ,
\end{eqnarray}
where the new 4th-rank Killing tensor has components
\begin{equation}\label{2KT}
K_{abcd} = Q^{\ix{(2)}} g^{}_{\ix{(ab}} g^{}_{\ix{cd)}} - 4 Q^{\ix{(2)}}_{\ix{(ab}}g^{}_{\ix{cd)}}
           + 2 Q^{}_{\ix{(ab}} Q^{}_{\ix{cd)}}
\end{equation}
and gives the $D$th constant of motion for $D=6$ and $D=7$.
 
Continuing in a similar fashion to get the $D$th constant of motion for $D=8$
and $D=9$,
\begin{eqnarray}\label{3KTconst}
C_3 &=& w^3{\rm Tr}(Q^3) - 6 w^2 {\rm Tr}(Q^3W) \nonumber\\
    &+& 6 w {\rm Tr}(Q^2W){\rm Tr}(QW) - 2[{\rm Tr}(QW)]^3\nonumber\\
    &=& (Q^{\ix{(3)}}g^{}_{\ix{ab}}g^{}_{\ix{cd}}g^{}_{\ix{cd}} - 6 Q^{\ix{(3)}}_{\ix{ab}}g^{}_{\ix{cd}}g^{}_{\ix{ef}} \nonumber\\
    &+& 6Q^{\ix{(2)}}_{\ix{ab}}Q^{\ix{(1)}}_{\ix{cd}}g^{}_{\ix{ef}}-2Q^{\ix{(1)}}_{\ix{ab}}Q^{\ix{(1)}}_{\ix{cd}}Q^{\ix{(1)}}_{\ix{ef}}) u^a u^b u^c u^d u^e u^f
    \nonumber\\
    &=& K_{abcdef} u^a u^b u^c u^d u^e u^f,
\end{eqnarray}
where the new $6$th-rank Killing tensor is
\begin{eqnarray}\label{3KT} 
K_{abcdef}&=&Q^{\ix{(3)}}g^{}_{\ix{(ab}}\,g^{}_{\ix{cd}}g^{}_{\ix{ef)}}
-6Q^{\ix{(3)}}_{\ix{(ab}}g^{}_{\ix{cd}}g^{}_{\ix{ef)}}\nonumber\\
&+& 6Q^{\ix{(2)}}_{\ix{(ab}}Q^{}_{\ix{cd}}g^{}_{\ix{ef)}}
-2Q^{}_{\ix{(ab}}Q^{}_{\ix{cd}}Q^{}_{\ix{ef)}}.                      
\end{eqnarray}

To finish the explicit expressions for all $D$ constants of motion (not
counting the constants from the metric and Killing vectors) up through $D=11$,
the highest dimension generally considered in superstring/M theory, we
calculate
\begin{eqnarray}\label{4KTconst}
C_4 &=& w^4{\rm Tr}(Q^4) - 8w^3{\rm Tr}(Q^4W) \nonumber\\
    &+& 8w^2{\rm Tr}(Q^3W){\rm Tr}(QW) + 4 w^2 [{\rm Tr}(Q^2W)]^2 \nonumber\\
    &-& 8 w {\rm Tr}(Q^2W)[{\rm Tr}(QW)]^2 + 2[{\rm Tr}(QW)]^4 \nonumber\\
&=&(Q^{\ix{(4)}}g_{ab}g_{cd}g_{ef}g_{gh}
  -8Q^{\ix{(4)}}_{\ix{ab}}g^{}_{\ix{cd}}g^{}_{\ix{ef}}g^{}_{\ix{gh}} \nonumber\\
&+&8Q^{\ix{(3)}}_{\ix{ab}}Q^{\ix{(1)}}_{\ix{cd}}g^{}_{\ix{ef}}g^{}_{\ix{gh}}
  +4Q^{\ix{(2)}}_{\ix{ab}}Q^{\ix{(2)}}_{\ix{cd}}g^{}_{\ix{ef}}g^{}_{\ix{gh}} \nonumber\\
&-&8Q^{\ix{(2)}}_{\ix{ab}}Q^{\ix{(1)}}_{\ix{cd}}Q^{\ix{(1)}}_{\ix{ef}}g^{}_{\ix{gh}}
   +2Q^{\ix{(1)}}_{\ix{ab}}Q^{\ix{(1)}}_{\ix{cd}}Q^{\ix{(1)}}_{\ix{ef}}Q^{\ix{(1)}}_{\ix{gh}})  \nonumber\\
&&\qquad \times u^a u^b u^c u^d u^e u^f u^g u^h \nonumber\\
&=&K_{abcdefgh} u^a u^b u^c u^d u^e u^f u^g u^h,
\end{eqnarray}
with the corresponding $8$th-rank Killing tensor being
\begin{eqnarray}\label{4KT}
&&K_{abcdefgh}=Q^{\ix{(4)}}g^{}_{\ix{(ab}}\,g^{}_{\ix{cd}}g^{}_{\ix{ef}}g^{}_{\ix{gh)}}
-8Q^{\ix{(4)}}_{\ix{(ab}}g^{}_{\ix{cd}}g^{}_{\ix{ef}}g^{}_{\ix{gh)}}\nonumber\\
&&\qquad+8Q^{\ix{(3)}}_{\ix{(ab}}Q^{}_{\ix{cd}}g^{}_{\ix{ef}}g^{}_{\ix{gh})}
+4Q^{\ix{(2)}}_{\ix{(ab}}Q^{\ix{(2)}}_{\ix{cd}}g^{}_{\ix{ef}}g^{}_{\ix{gh})}\nonumber\\
&&\qquad-8Q^{\ix{(2)}}_{\ix{(ab}}Q^{}_{\ix{cd}}Q^{}_{\ix{ef}}g^{}_{\ix{gh})}+2Q^{}_{\ix{(ab}}Q^{}_{\ix{cd}}Q^{}_{\ix{ef}}Q^{}_{\ix{gh)}}.
\end{eqnarray}

For a spacetime with $D=2n+\varepsilon$ dimensions, we get $D-n=n+\varepsilon$
constants of motion from the $D-n$ Killing vectors and one constant of motion,
$w = \mathbf{u}\cdot\mathbf{u}$, from the metric Killing tensor.  Therefore, we
need $C_j$ up through $j=n-1$ to give the remainder of the $D$ constants of
motion.

We can explicitly show \cite{KKPV} that all of these $D$ constants of motion
are independent of each other, as functions of the velocity components $u^a$,
by calculating the Jacobian of this transformation.  When all of the first $n$
velocity components are nonzero, and at a generic point of the manifold where
none of the $(x_\mu)^2$'s coincide (which would actually give a coordinate
singularity if any did coincide), we find that the Jacobian is nonzero.  Key to
the proof is the fact that in the constant $C_j$, the coefficient of ${\rm
Tr}(Q^jW)$, given explicitly in Eq. (\ref{cj}), is nonzero, as well as the fact
that the $n$ pairs of eigenvalues of the matrix $Q$, namely the $(x_\mu)^2$'s,
are all different when none of the $(x_\mu)^2$'s coincide.  Therefore, the
${\rm Tr}(Q^jW)$ term for each higher $j$ up to $n-1$ gives a function of the
velocity components that is independent of any of the terms with lower $j$.  We
can also prove \cite{KKPV} that the Poisson bracket between any pair of these
$D$ constant vanishes, so these constants are in involution, a sufficient
condition for the integrable motion to be completely integrable (see, e.g.,
\cite{Arn,Koz}).

In summary, we have shown that geodesic motion is completely integrable for all
Kerr-NUT-AdS metrics \cite{CLP} in all dimensions and with arbitrary rotation
and NUT parameters.  However, this has not enabled us (at least yet) to
separate the Hamilton-Jacobi, Dirac, and Klein-Gordon equations.

\noindent

\section*{Acknowledgments}

D.N.P. thanks the Natural Sciences and Engineering Research Council of Canada
for financial support.  D.K. is grateful to the Golden Bell Jar Graduate
Scholarship in Physics at the University of Alberta.  P.K. appreciates the
hospitality of the University of Alberta.  We thank Gary Gibbons for asking us
whether the Poisson brackets of pairs of our constants commute and Paul Davis
for informing us that he has independently shown that all odd-dimensional
Kerr-AdS black holes with three arbitrary rotation parameters are Liouville
integrable.  We have also been aided by discussions with Hari Kunduri, Dmitri
Pogosyan, and Chris Pope.


\end{document}